# Activation cross sections of deuteron induced reactions on silver in the 33-50 MeV energy range

F. Ditrói[a,1], F. Tárkányi[a], S. Takács[a], A. Hermanne[b], A.V. Ignatyuk[c]

a Institute for Nuclear Research, Hungarian Academy of Sciences (ATOMKI), 4026 Debrecen, Hungary

b Cyclotron Laboratory, Vrije Universiteit Brussel (VUB), Laarbeeklaan 103, 1090 Brussels, Belgium

c Institute of Physics and Power Engineering (IPPE), Obninsk 249020, Russia

*Abstract*

Excitation functions were measured for the $^{nat}Ag(d,x)^{105,104}Cd$, $^{110m,108m,106m,105g,104g}Ag$ and $^{101}Pd$, $^{105,101m}Rh$ reactions over the energy range 33–50 MeV by using the stacked foil activation technique and subsequent high-resolution gamma-spectrometry. We present the first experimental cross section data above 40 MeV for all of these reactions and the first experimental cross section data for $^{nat}Ag(d,x)^{108m,104g}Ag$ and $^{105,103}Rh$. The experimental data are compared with results of the model calculations performed with the ALICE-D, EMPIRE-D theoretical nuclear reaction model codes and with the TALYS code results as available in the TENDL-2014 and -2015 on-line libraries.

Keywords: silver target; deuteron activation; cadmium, silver, palladium and rhodium radioisotopes; physical yield; Thin Layer Activation

---

[1] Corresponding author: ditroi@atomki.hu

## 1. Introduction

The study of the $^{nat}$Ag(d,x) reactions was initiated by several motives: Silver is an important material for the accelerator technology. Activation cross sections of proton and deuteron induced reactions are included in the FENDL library (IAEA, 2011). The $^{107}$Cd, $^{109}$Cd and $^{110m}$Ag nuclides have large importance in medical and industrial applications (Ditrói et al., 1997; Gruverman and Kruger, 1959; Long et al., 1991; Uddin et al., 2006). In particular, the short-lived isotopes $^{107m}$Ag ($T_{1/2}$ =44.3 s) and $^{109m}$Ag ($T_{1/2}$= 39.6 s) adopted for nuclear medicine, are provided through the $^{107}$Cd -$^{107m}$Ag and $^{109}$Cd - $^{109m}$Ag generator systems. The results and the production routes for both of these isotopes will be published in a separate paper. The $^{105g}$Ag and $^{110m}$Ag have decay characteristics that allow interesting use in Thin Layer Activation (TLA) (Ditrói et al., 1997; Ditrói et al., 2012).

In the frame of our systematic study of the experimental data base of light charged particle induced reactions on metals we studied earlier reactions induced on silver targets by protons up to 70 MeV (Uddin et al., 2005), by deuterons up to 40 MeV (Uddin et al., 2006) and by alpha particles up to 40 MeV (Takács et al., 2010). A literature search showed that for deuteron bombardment of silver investigations were done only up to 40 MeV incident energy by several authors (Baron and Cohen, 1963; Dmitriev et al., 1967; Gruverman and Kruger, 1959; Lalli et al., 1976; Long et al., 1991; Ornstein et al., 1953; Peng et al., 1992; Röhm et al., 1970; Weixiang et al., 1989).

Having the possibility to get higher energy deuteron beams we extended the measurements of the excitation functions of $^{nat}$Ag+d reactions up to 50 MeV incident energy in this work.

## 2. Experiment and data evaluation

The cross section measurements were performed using the activation method combining a stacked foil irradiation and high resolution gamma-ray spectrometry. Cross section data were deduced relative to the re-measured excitation functions of monitor reactions.
The stack was irradiated at an external beam line of the Cyclone 90 cyclotron of the Université Catholique in Louvain la Neuve (LLN) for 40 min with a 50 MeV, 50 nA deuteron beam. The stack contained a sequence of 10 blocks of Al (10 μm), Sr(NO$_3$)$_2$ (3 μm), Al (50 μm), Er (32 μm), Al (10 μm), Ba(NO$_2$) (2 μm, sedimented), Al (50 μm) and Ag (10 μm) foils. The 10 Ag targets covered the 50-33 MeV energy range.

The targets were irradiated in a short Faraday cup. Gamma-ray spectra were measured with HPGe detectors coupled to Canberra multi-channel analyzers equipped with GENIE acquisition software. Four series of gamma-ray spectra were measured to follow the decay and started at different times after the end of bombardment: 8.6 - 10.2 h, 51.7 - 54.7 h, 174.7 - 191.7 h and 885.9 - 959.7 h, respectively

The gamma-spectra were evaluated by the automatic fitting algorithm included in the Genie 2000 package or in an iterative process using the Forgamma (Canberra, 2000; Székely, 1985) codes. The used decay data taken from NUDAT2.6 (NuDat, 2014) and the Q values of the contributing reactions (Q value calculator (Pritychenko and Sonzogni, 2003)) are shown in Table 1. The recommended data for simultaneously measured excitation functions of the $^{27}$Al(d,x)$^{22,24}$Na monitor reactions were taken from the IAEA database (Tárkányi et al., 2001). As Ag is not monoisotopic ($^{107}$Ag: 51.83%, $^{109}$Ag: 48.17 % natural abundance), so called elemental cross sections were determined. Uncertainties of cross sections were determined according to the recommendation in (International-Bureau-of-Weights-and-Measures, 1993) by taking the square root of the sum of the squared errors of all individual contributions: beam current (7%), beam-loss corrections (max. of 1.5%), target thickness (1%), detector efficiency (5%), photo peak area determination and counting statistics (1-20 %).

The beam energies in the targets were initially (from stack preparation) obtained from an energy loss calculation based on the incident energy and the Andersen's (Andersen and Ziegler, 1977) polynomial approximation of stopping powers. If needed, energy and beam current corrections were applied based on the results of the fitted monitor reactions (final) (Tárkányi et al., 1991). Uncertainties of energies were estimated by following the cumulative effects during the energy degradation taking into account the possible uncertainties (primary energy, target thickness, energy straggling, correction to monitor reaction).

Table 1 Decay data of the investigated reaction products (NuDat, 2014; Pritychenko and Sonzogni, 2003)

| Nuclide Decay path Level energy | Half-life | $E_\gamma$(keV) | $I_\gamma$(%) | Contributing reaction | Q-value (MeV) **GS-GS** |
|---|---|---|---|---|---|
| $^{105}$Cd<br>ε: 100 % | 55.5 min | 346.9<br>433.2<br>648.5<br>961.8 | 4.2<br>2.8<br>1.6<br>4.7 | $^{107}$Ag(d,4n)<br>$^{109}$Ag(d,6n) | -23.22<br>-39.68 |
| $^{104}$Cd<br>ε: 100 % | 57.7 min | 559.0<br>709.3 | 6.3<br>19.5 | $^{107}$Ag(d,5n)<br>$^{109}$Ag(d,6n) | -31.66<br>-39.68 |
| $^{110m}$Ag<br>IT: 1.33 %<br>β⁻: 98.67 %<br>117.595 keV | 249.83 d | 657.8<br>677.6<br>706.7<br>763.9<br>884.7<br>937.5<br>1384.3<br>1505.0 | 95.6<br>10.7<br>16.7<br>22.6<br>75.0<br>35.0<br>25.1<br>13.3 | $^{109}$Ag(d,p) | -4.58 |
| $^{108m}$Ag | 438 a | 433.9<br>614.3<br>722.9 | 90.5<br>89.8<br>90.8 | $^{107}$Ag(d,p)<br>$^{109}$Ag(d,p2n) | 5.05<br>-11.41 |
| $^{106m}$Ag<br>ε: 100 %<br>89.667 keV | 8.28 d | 406.2<br>429.6<br>451.0<br>616.2<br>717.3<br>748.4<br>804.3<br>824.7 | 13.4<br>13.2<br>28.2<br>21.6<br>28.9<br>20.6<br>12.4<br>15.3 | $^{107}$Ag(d,p2n)<br>$^{109}$Ag(d,p4n) | -11.76<br>-28.22 |
| $^{105g}$Ag<br>ε: 100 % | 41.29 d | 280.4<br>344.5<br>443.4<br>644.6 | 30.2<br>41.4<br>10.5<br>11.1 | $^{107}$Ag(d,p3n)<br>$^{109}$Ag(d,p5n)<br>$^{105}$Cd decay | -19.70<br>-36.16 |
| $^{104m}$Ag<br>IT: 0.07 %<br>ε: 99.93 %<br>6.9022 keV | 33.5 min | 996.1<br>1238.8 | 0.50<br>3.9 | $^{107}$Ag(d,p4n)<br>$^{109}$Ag(d,p6n)<br>$^{104}$Cd decay | -29.73<br>-46.18 |
| $^{104g}$Ag<br>ε: 100 % | 69.2 min | 767.6<br>857.9<br>941.6 | 65.7<br>10.45<br>25.0 | $^{107}$Ag(d,p4n)<br>$^{109}$Ag(d,p6n) | -29.73<br>-46.18 |
| $^{101}$Pd<br>ε: 100 % | 8.47 h | 269.7<br>296.3<br>566.0<br>590.4 | 6.4<br>19.0<br>3.4<br>12.1 | $^{107}$Ag(d,2p6n)<br>$^{109}$Ag(d,2p8n) | -52.85<br>-70.58 |
| $^{105}$Rh<br>β⁻: 100 % | 35.36 h | 306.1<br>318.9 | 5.1<br>19.1 | $^{107}$Ag(d,3pn)<br>$^{109}$Ag(d,3p3n) | -17.36<br>-33.81 |
| $^{101m}$Rh<br>IT: 7.20 %<br>ε :92.8 %<br>157.41 keV | 4.34 d | 306.9<br>545.1 | 81<br>4.3 | $^{107}$Ag(d,3p5n)<br>$^{109}$Ag(d,3p7n)<br>Decay of<br>$^{101}$Ag-$^{101}$Pd | -50.08<br>-66.54 |

Increase the Q-values if compound particles are emitted by: np-d, +2.2 MeV; 2np-t, +8.48 MeV; n2p-$^3$He, +7.72 MeV; 2n2p-α, +28.30 MeV.
Decrease Q-values for isomeric states with level energy of the isomer

## 3. Model calculations

The cross sections of the investigated reactions were calculated using the pre-compound model codes ALICE-IPPE (Dityuk et al., 1998) and EMPIRE-II (Herman et al., 2007) modified for deuterons by Ignatyuk (D versions) (Ignatyuk, 2010, 2011). Independent data for metastable states from the ALICE-D code were obtained by applying the isomeric ratios calculated with EMPIRE-D to the total cross sections from ALICE-D. The experimental data are also compared with the cross section data reported in the TENDL-2014 (Koning et al., 2014) and TENDL-2015 (Koning et al., 2015) nuclear reaction data libraries. The TENDL libraries are based on default and adjusted TALYS (1.6) calculations (Koning and Rochman, 2012).

## 4. Results

### 4.1 Cross sections

The experimental cross section data are shown graphically in Figs. 1-10 in comparison with earlier published values and with the predictions of the theoretical codes. The numerical data are given in Table 2.

### 4.1.1 $^{nat}Ag(d,x)^{105}Cd$

Due to the long cooling time and low gamma-ray intensities $^{105}$Cd ($T_{1/2}$ = 55.5 min, ε: 100 %) was identified with poor statistics only in a few targets (Fig. 1). No earlier published experimental data were found for comparison. The experimental data are lower than the theoretical results that show large mutual differences in shape and amplitude.

### 4.1.2 $^{nat}Ag(d,x)^{104}Cd$

In spite of the short half-life of $^{104}$Cd ($T_{1/2}$ =57.7 min, ε: 100 %) compared to the used cooling time we could deduce cross section data for production of $^{104}$Cd from the first measurements after EOB (Fig. 2). No earlier experimental data are available in the literature. The agreement with the two TENDL data sets is acceptable (a small energy shift is noted, 20% difference between the sets). The ALICE-D predictions significantly overestimate the experimental data. There is good agreement with the predictions of EMPIRE-D in the overlapping energy range.

### 4.1.3 $^{nat}Ag(d,x)^{110m}Ag$

The radionuclide $^{110}$Ag has a short-lived ground-state ($T_{1/2}$=24.5 s) and a long-lived excited isomeric state $^{110m}$Ag ($T_{1/2}$= 249.8 d, IT: 1.33 %, β$^-$: 98.67 %) that decays by internal transition to $^{110g}$Ag and by β$^-$ decay to stable $^{110}$Cd. We could only measure the formation of the long-lived isomer (Fig. 3). The experimental data show good agreement with literature values in the overlapping energy region (Uddin et al., 2006). The predictions of the model codes are very diverging and the agreement with the experimental data is poor. Our new experimental results are between the EMPIRE-D (overestimates) and ALICE-D (partly underestimates) predictions. The TENDL versions give much lower values. The best approximation comes from the systematics taken from the experimental results from of similar reactions in the mass region.

### 4.1.4 $^{nat}Ag(d,x)^{108m}Ag$

The radionuclide $^{108g}$Ag has a short-lived ground-state ($T_{1/2}$ = 2.382 min, β$^-$: 97.15 %, ε: 2.85 %) and a long-lived isomeric state $^{108m}$Ag ($T_{1/2}$= 438 a, IT: 8.7 %, ε: 91.3 %). We could only measure the production of the long-lived isomer (Fig. 4). No earlier experimental data were found in the literature. The theoretical predictions are very diverging.

### 4.1.5 $^{nat}Ag(d,x)^{106m}Ag$

The radionuclide $^{106}$Ag has a short-lived ground-state ($T_{1/2}$ = 23.96 min) and an excited isomeric state $^{106m}$Ag ($T_{1/2}$ = 8.28 d, EC: 100%). Only results for the longer-lived isomeric state were obtained (Fig. 5). The new values are somewhat higher than Uddin et al. 2006 (Uddin et al., 2006). The agreement with the ALICE-D and EMPIRE-D predictions is acceptable. The TENDL libraries significantly overestimate the experimental results.

### 4.1.6 $^{nat}Ag(d,x)^{105g}Ag$

The radionuclide $^{105}$Ag has a ground-state ($T_{1/2}$ = 41.29 d) and a short-lived excited isomeric state $^{105m}$Ag ($T_{1/2}$ = 7.23 min, IT: 99.66%). We could measure the full cumulative production of $^{105g}$Ag (Fig. 6) containing the contribution from the decay of the isomer, and from the $^{105}$Cd ($T_{1/2}$ = 55.5 min, ε: 100 %) parent. Good agreement with the earlier experimental data was found. The predicted values for cumulative formation in the two TENDL libraries describe rather well the experimental behavior. The results for ALICE-D and EMPIRE-D are too peaked and too high.

### 4.1.7 $^{nat}Ag(d,x)^{104g}Ag$

The radionuclide $^{104}$Ag also has a short-lived ground-state ($T_{1/2}$ = 69.2 min), and a shorter-lived excited isomeric state $^{104m}$Ag ($T_{1/2}$= 33.5 min, IT: 0.07 %). In the first series of spectra we could not detect independent gamma-lines from decay of $^{104m}$Ag and the stronger 550 keV common gamma-line gives the same cross sections as the cross sections obtained from independent gamma-lines of $^{104g}$Ag. The measured cross section for the $^{104g}$Ag is hence cumulative (m+) but the contribution from the $^{104m}$Ag is negligible (Fig. 7). The theoretical descriptions are acceptable, except the EMPIRE-D, which gives somewhat lower results.

### 4.1.8 $^{nat}Ag(d,x)^{101}Pd$

The radionuclide $^{101}$Pd (T$_{1/2}$ = 8.47 h, ε: 100 %) is produced directly and through the decay of $^{101g}$Ag. The measured production cross sections of $^{101}$Pd are cumulative, obtained from the spectra measured after 'complete' decay of $^{101}$Ag ($^{101g}$Ag T$_{1/2}$ = 11.1 min, EC: 100%, including $^{101m}$Ag (T$_{1/2}$ = 3.1 s, IT 100%) (Fig. 8). No earlier experimental data were found in the literature. The threshold of the theoretical predictions is lower and the different codes behave in a quite different way above 50 MeV.

### 4.1.9 $^{nat}Ag(d,x)^{105}Rh$

The radionuclide $^{105}$Rh has a long-lived ground-state $^{105g}$Rh (T$_{1/2}$ = 35.36 h, β$^-$: 100%) and a short-lived metastable state $^{105m}$Rh (T$_{1/2}$ = 40 s, IT: 100%). Our results, measured after the complete decay of the isomeric state, present cumulative cross sections (Fig. 9). The $^{105}$Rh is produced directly (if we neglect the small contribution from the decay of the $^{105}$Ru (T$_{1/2}$ = 4.44 h) isotope). The obtained experimental data show the best agreement with the ALICE-D results.

### 4.1.10 $^{nat}Ag(d,x)^{101m}Rh$

Out of the two levels of $^{101}$Rh ($^{101g}$Rh - T$_{1/2}$ = 3.3 a, $^{101m}$Rh - T$_{1/2}$ = 4.34 d, IT: 7.20 % ε :92.8 %) we could identify only the gamma-lines of the isomeric state. The radionuclide $^{101m}$Rh is produced directly and in principle can also be produced by the decay of $^{101}$Pd (T$_{1/2}$ = 8.47 h, ε: 100 %) that decays also to the isomeric state of $^{101}$Rh. The cross sections were determined from spectra taken after the "complete" decay of $^{101}$Pd, in such a way they are cumulative (Fig. 10). The experimental data are higher comparing to the theory and the effective threshold is also shifted, probably because of the presence of complex particle emission (see Table 1), which was not handled correctly by the model codes.

Table 2  Experimental cross sections of $^{105,104}$Cd, $^{110m,108m,106m,105,104g}$Ag $^{101}$Pd and $^{105,101m}$Rh nuclear reactions

| Energy E + ΔE (MeV) | | $^{105}$Cd | | $^{104}$Cd | | $^{110m}$Ag | | $^{108m}$Ag | | $^{106m}$Ag | | $^{105g}$Ag | | $^{104g}$Ag | | $^{101}$Pd | | $^{105}$Rh | | $^{101m}$Rh | |
|---|---|---|---|---|---|---|---|---|---|---|---|---|---|---|---|---|---|---|---|---|---|
| | | Cross section σ + Δσ (mb) | | | | | | | | | | | | | | | | | | | |
| 48.8 | 0.3 | 35.3 | 29.6 | 35.6 | 5.3 | 2.8 | 0.4 | | | 142.4 | 16.0 | 353.8 | 39.8 | 137.9 | 15.6 | 9.5 | 1.1 | 5.1 | 0.7 | 18.4 | 2.1 |
| 47.3 | 0.3 | | | 33.2 | 4.7 | 3.1 | 0.4 | 86.2 | 13.9 | 134.0 | 15.1 | 363.3 | 40.8 | 101.2 | 11.5 | 5.1 | 0.6 | 5.4 | 0.7 | 10.3 | 1.2 |
| 45.7 | 0.4 | | | 20.8 | 4.6 | 3.5 | 0.4 | 110.4 | 16.5 | 123.4 | 13.9 | 396.8 | 44.6 | 66.9 | 7.7 | 2.4 | 0.3 | 4.5 | 0.5 | 5.3 | 0.6 |
| 44.1 | 0.5 | 111.0 | 111.7 | 7.7 | 2.4 | 3.3 | 0.4 | 114.4 | 23.5 | 113.5 | 12.8 | 383.6 | 43.1 | 35.8 | 4.2 | 0.9 | 0.2 | 4.3 | 0.5 | 2.5 | 0.3 |
| 42.4 | 0.5 | 55.4 | 29.1 | | | 4.1 | 0.5 | 97.4 | 15.2 | 115.4 | 12.9 | 393.9 | 44.3 | 19.1 | 2.4 | | | 4.3 | 0.5 | 1.6 | 0.2 |
| 40.7 | 0.6 | | | | | 4.5 | 0.5 | 107.1 | 16.2 | 117.3 | 13.2 | 374.9 | 42.1 | 5.2 | 1.3 | | | 3.7 | 0.4 | 1.3 | 0.2 |
| 38.9 | 0.6 | | | | | 4.0 | 0.5 | 120.1 | 17.6 | 113.9 | 12.8 | 313.3 | 35.2 | 3.0 | 1.0 | | | 3.5 | 0.4 | 0.9 | 0.2 |
| 37.1 | 0.7 | 22.0 | 49.0 | | | 4.7 | 0.6 | 107.7 | 15.6 | 117.3 | 13.2 | 254.1 | 28.6 | | | | | 2.8 | 0.4 | 0.8 | 0.2 |
| 35.2 | 0.7 | | | | | 4.8 | 0.5 | 101.4 | 14.6 | 116.5 | 13.1 | 176.7 | 19.9 | | | | | 1.9 | 0.3 | 0.5 | 0.1 |
| 33.2 | 0.8 | 71.3 | 25.8 | | | 5.7 | 0.6 | 95.0 | 16.7 | 115.9 | 13.0 | 94.5 | 10.6 | | | | | 1.3 | 0.2 | 0.3 | 0.05 |

## 5. Yields

Based on a spline fit to our experimental data, integral thick target yields were deduced for radioisotopes, where the data has good statistics and the cross section could be extrapolated or extended down to the threshold of the particular reactions. In the case $^{110m}$Ag and $^{106m}$Ag the previous experimental data of Uddin et al. (Uddin et al., 2006) were used to estimate the full cross section curve (Fig. 11). The yields are so called physical yields, calculated for an instantaneous irradiation (Bonardi, 1987; Otuka and Takács, 2015). The deduced yields are compared with the experimental thick target yields in the literature (Dmitriev et al., 1982), which was only available for $^{108m}$Ag, where the data of (Dmitriev et al., 1982) for the single 22 MeV point is below our curve in spite of the fact that they used $T_{1/2}$ = 127 years half-life instead of the new value of 438 years.

## 6. Thin layer activation

Some of the produced isotopes are suitable also for application in radioisotope tracing i.e. for wear measurement with Thin Layer Activation (TLA) (Ditrói et al., 1997; Ditrói et al., 2012). The $^{110m}$Ag ($T_{1/2}$ = 249.83 d) has conveniently long half-life and properly strong gamma-energies as well as it can be produced even with small compact cyclotrons or other low energy accelerators. The half-life of $^{105g}$Ag ($T_{1/2}$ = 41.29 d) enables quick wear measurement. Its gamma-energies are weaker but still proper for TLA and the production cross section is high, but only at higher energies, requiring a bigger accelerator. The calculated wear curves (activation profiles) are shown in Fig. 12. All calculations were made for the time of the end of the bombardment (EOB) with 1 µA beam currents and 1.5 h irradiation times. In the case of $^{110m}$Ag the optimum irradiation energy for constant activity distribution near to the surface is 11.9 MeV. The cross section data were taken from our previous paper (Uddin et al., 2006). In Fig. 12 the activation profiles achieved by an irradiation angle of 15° as well as activation profiles by perpendicular irradiation are shown. The constant activity region (within 1 %) is 10.1 and 39 µm for the angular and perpendicular irradiations, while the total activation depths are 50 and 194 µm, respectively. It means that according to the requirements of the particular tribological tasks the produced activation depth and specific activity can be varied in a wide range. For linear activity distribution only 8.1 MeV bombarding energy is required and the resulted activity distributions are also shown in Fig. 12. In this case the linearity regions are 4 and 15.4 µm (within 1 %) and the total activation depths are 25 and 96 µm, respectively. The same calculations were performed for the other TLA candidate $^{105g}$Ag in order to demonstrate the capability of TLA in the energy range of the present work. The beam currents and the irradiation times are the same as before, the results are shown in Fig. 13. It is obviously seen that the produced specific activities are 2 orders of magnitudes higher, but the activation requires a high energy accelerator. Because of the higher bombarding energies, the activation depths are also higher. The constant/linearity ranges and the total activation depths are 52.9, 204.4, 39.4 and 152.4 µm, as well as 334, 1291, 190 and 733 µm for the four calculated cases, respectively (see Fig. 13). It must also be taken into account that all curves were calculated for EOB, and in the case of practical application the activity of $^{105g}$Ag decreases much quicker than that of $^{110m}$Ag because of its shorter half-life. The radioisotope $^{105g}$Ag might further broaden the toolbox of the TLA method for wear measurements.

## 7. Summary and conclusions

In the frame of a systematic investigation of excitation functions of deuteron induced activation products, cross sections on silver were investigated in the 33-50 MeV deuteron energy range for applied purposes. The goal of this study was to improve the theoretical predictions too. Independent or cumulative cross sections for the production of $^{105,104}$Cd, $^{110m,108m,106m,105g,104g}$Ag $^{101}$Pd and $^{105,101m}$Rh were measured for the first time above 40 MeV. The new experimental data fit well to the earlier lower energy literature values. A comparison is made with the theoretical predictions obtained by our EMPIRE-D and ALICE-IPPE-D code calculations and with the TALYS data in the latest TENDL libraries. The results of the theoretical codes differ significantly, the agreement with the experimental data is often poor. Yield curves were also calculated for the isotopes having suitable data sets and compared with the single literature value, which differs from our results. Activity distribution curves were also deduced for the most proper TLA candidates in order to demonstrate the applicability of them in wear measurements.

*Acknowledgements*

This work was done in the frame of MTA-FWO (Vlaanderen) research projects. The authors acknowledge the support of research projects and of their respective institutions in providing the materials and the facilities for this work. We thank the crew of the LLN cyclotron for performing numerous irradiations for us.

**Figures**

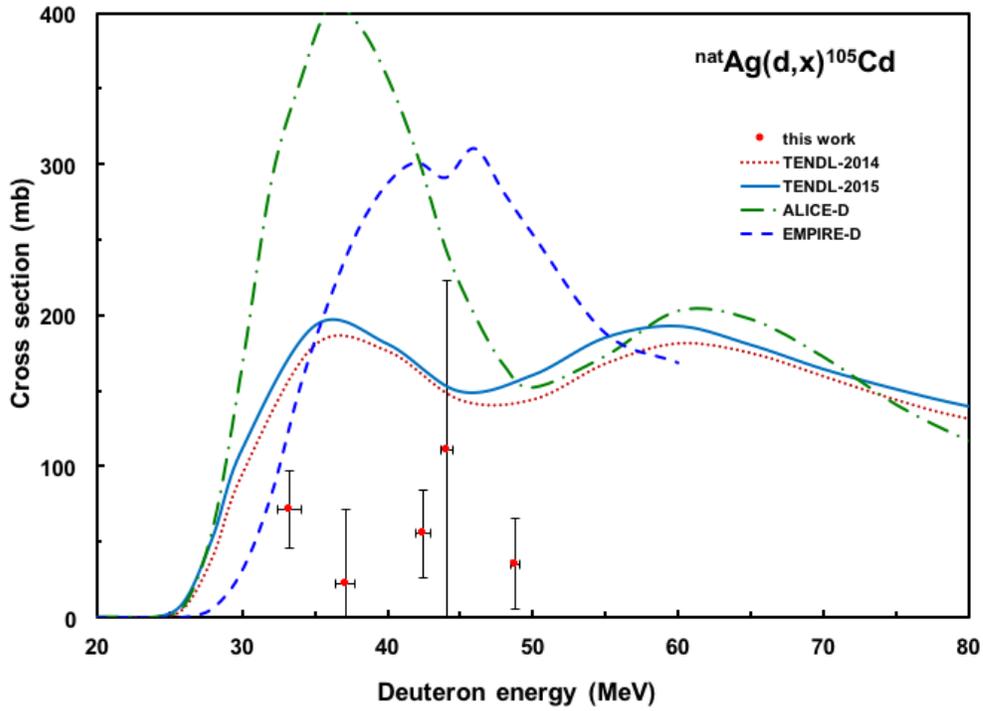

Fig.1. Excitation function of the $^{nat}Ag(d,x)^{105}Cd$ reaction

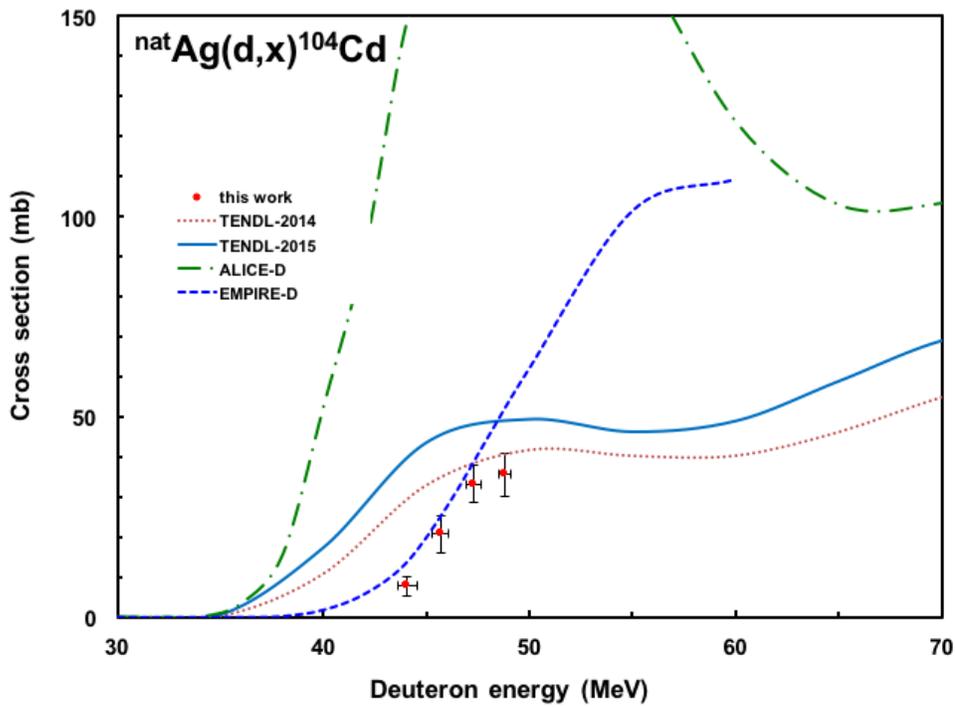

Fig. 2. Excitation function of the $^{nat}Ag(d,x)^{104}Cd$ reaction

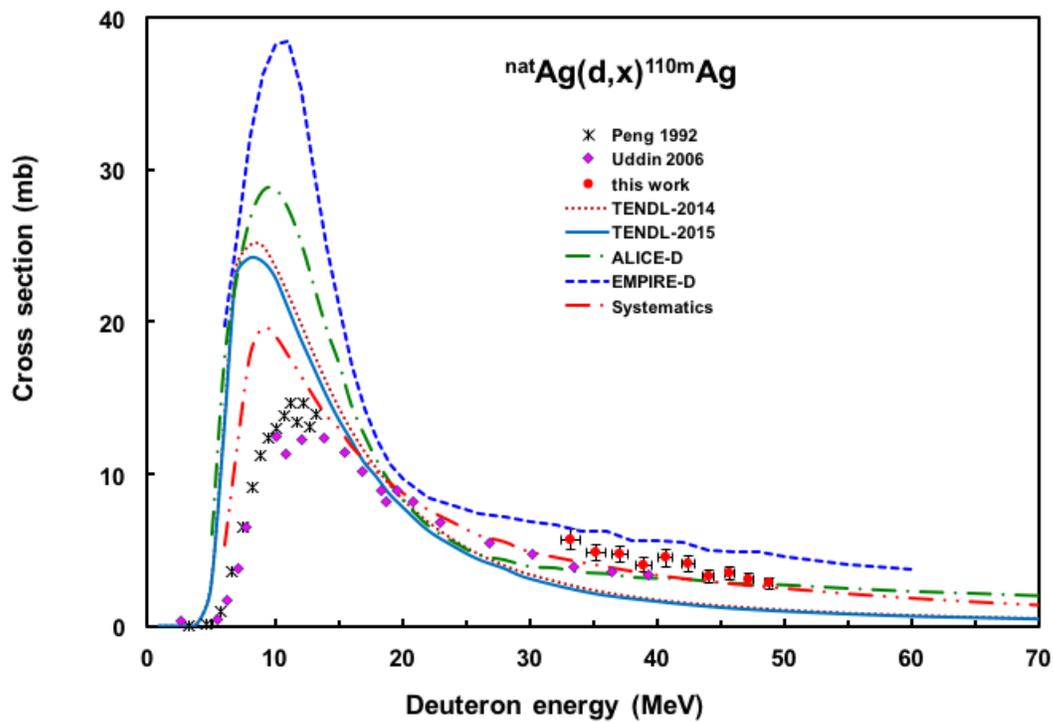

Fig. 3. Excitation function of the $^{nat}Ag(d,x)^{110m}Ag$ reaction

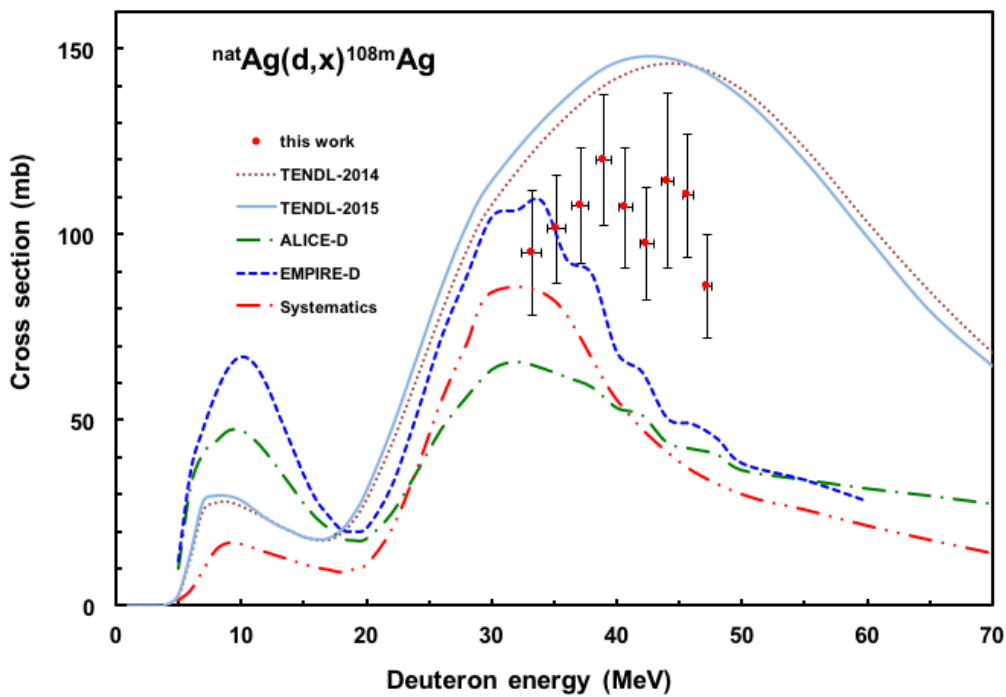

Fig. 4. Excitation function of the $^{nat}Ag(d,x)^{108m}Ag$ reaction

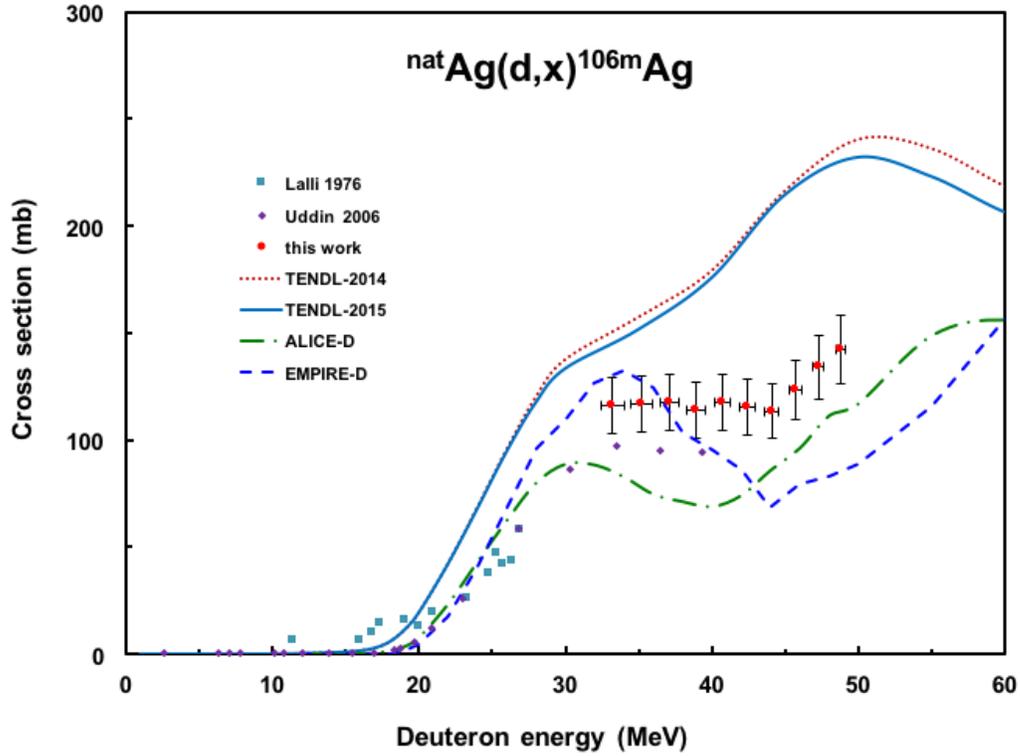

Fig. 5. Excitation function of the $^{nat}$Ag(d,x)$^{106m}$Ag reaction

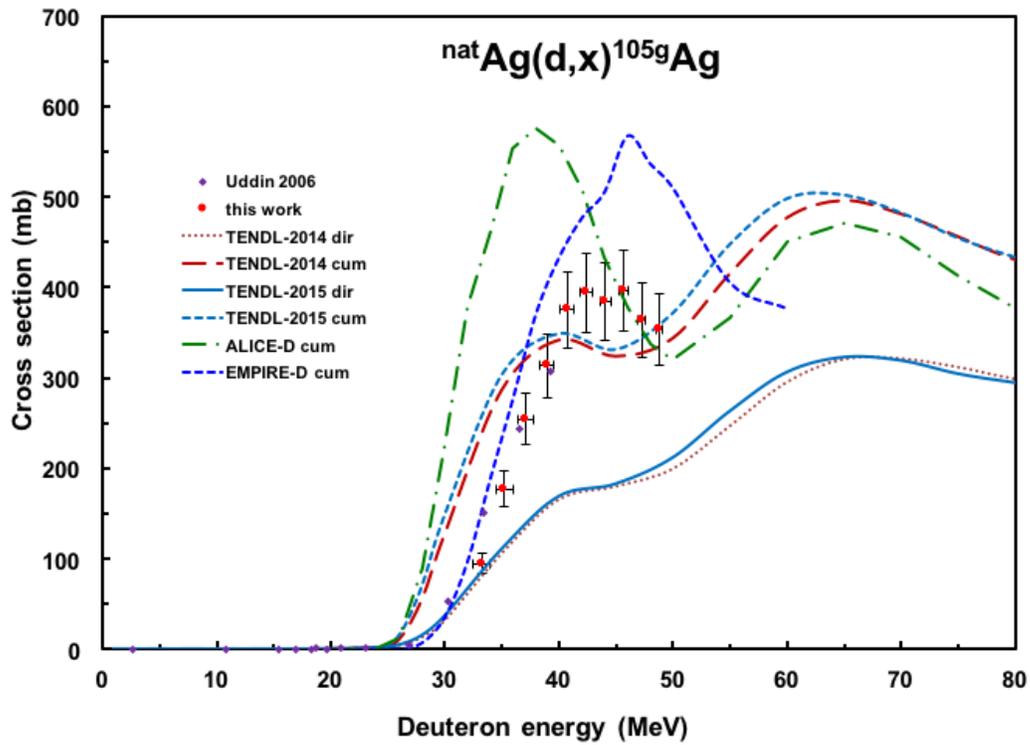

Fig. 6. Excitation function of the $^{nat}$Ag(d,x)$^{105g}$Ag reaction

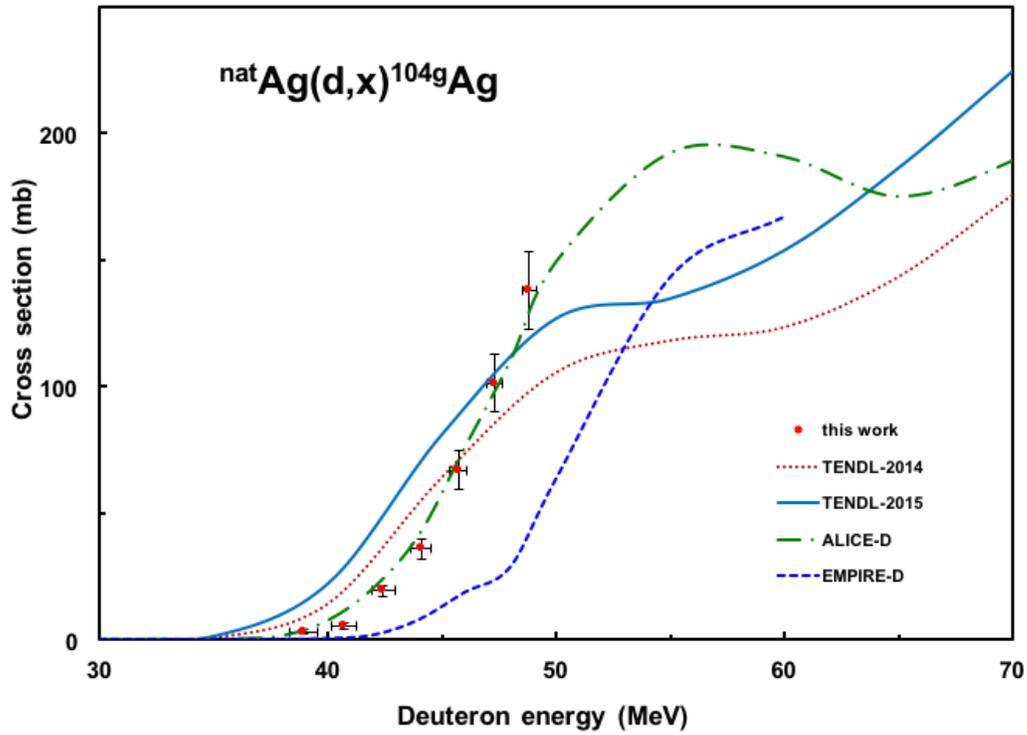

Fig. 7. Excitation function of the $^{nat}Ag(d,x)^{104g}Ag$ reaction

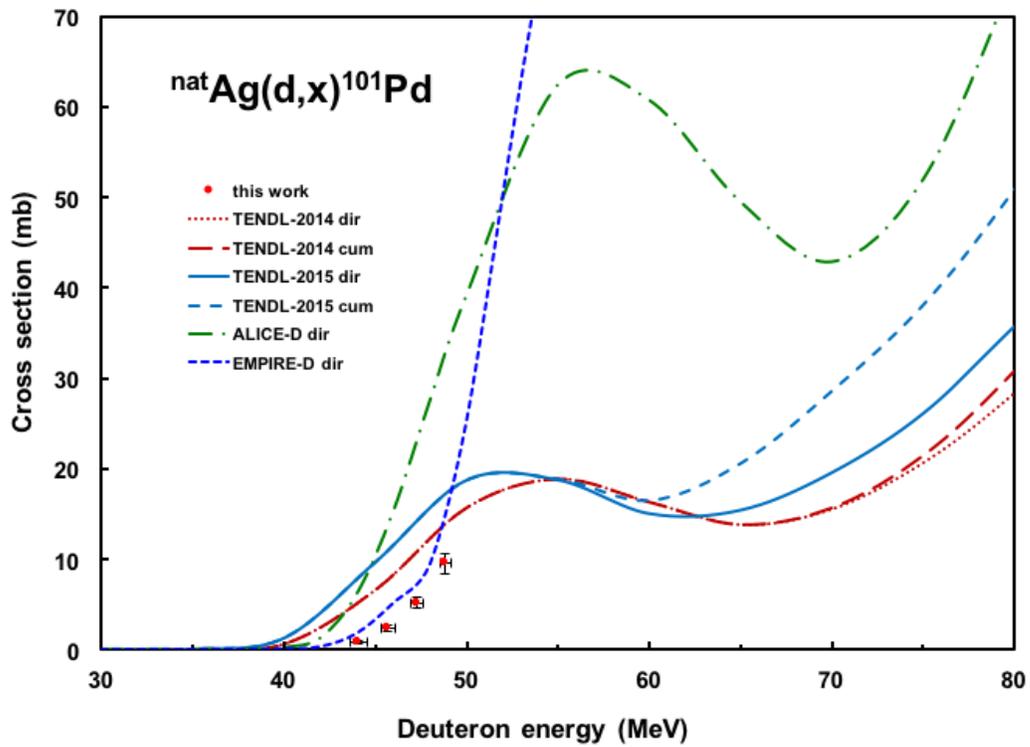

Fig. 8. Excitation function of the $^{nat}Ag(d,x)^{101}Pd$ reaction

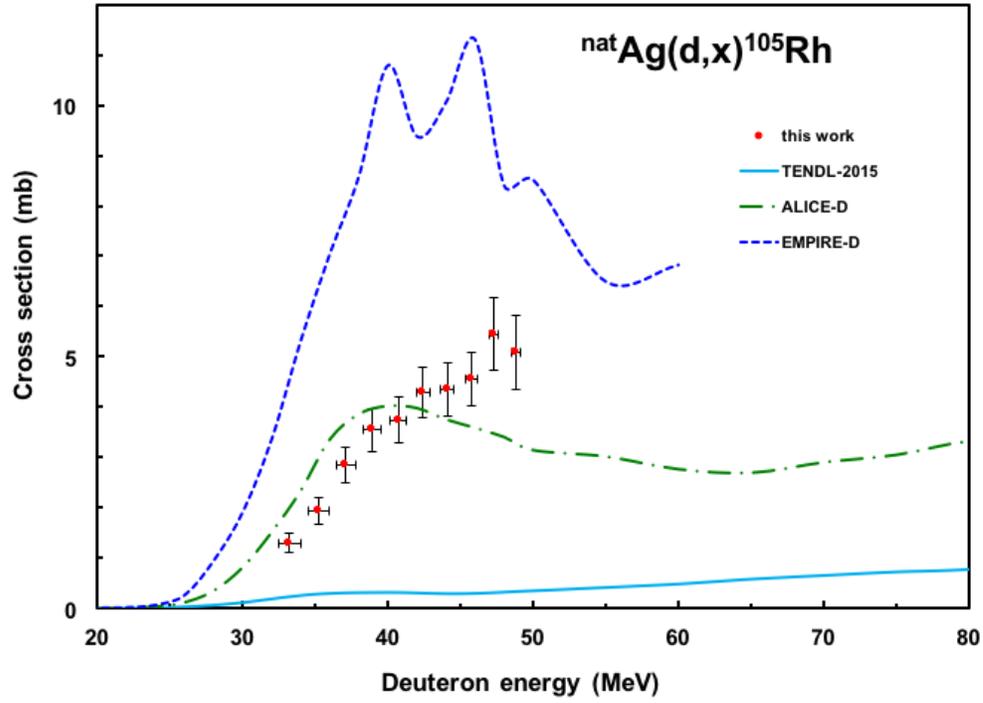

Fig. 9. Excitation function of the $^{nat}Ag(d,x)^{105}Rh$ reaction

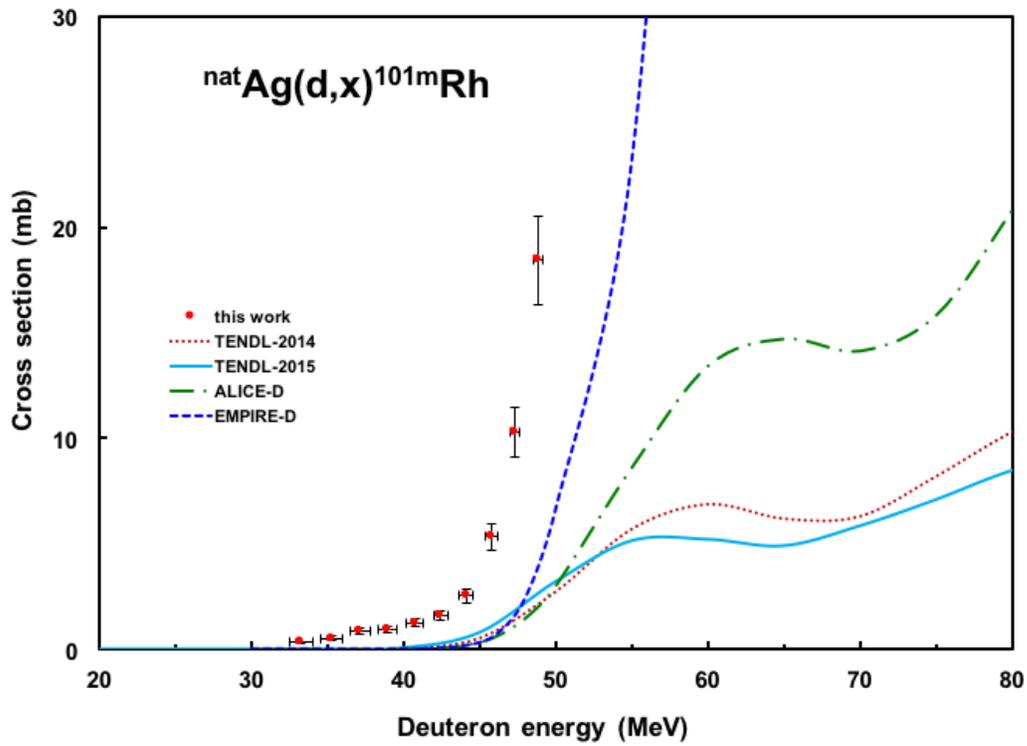

Fig. 10. Excitation function of the $^{nat}Ag(d,x)^{101m}Rh$ reaction

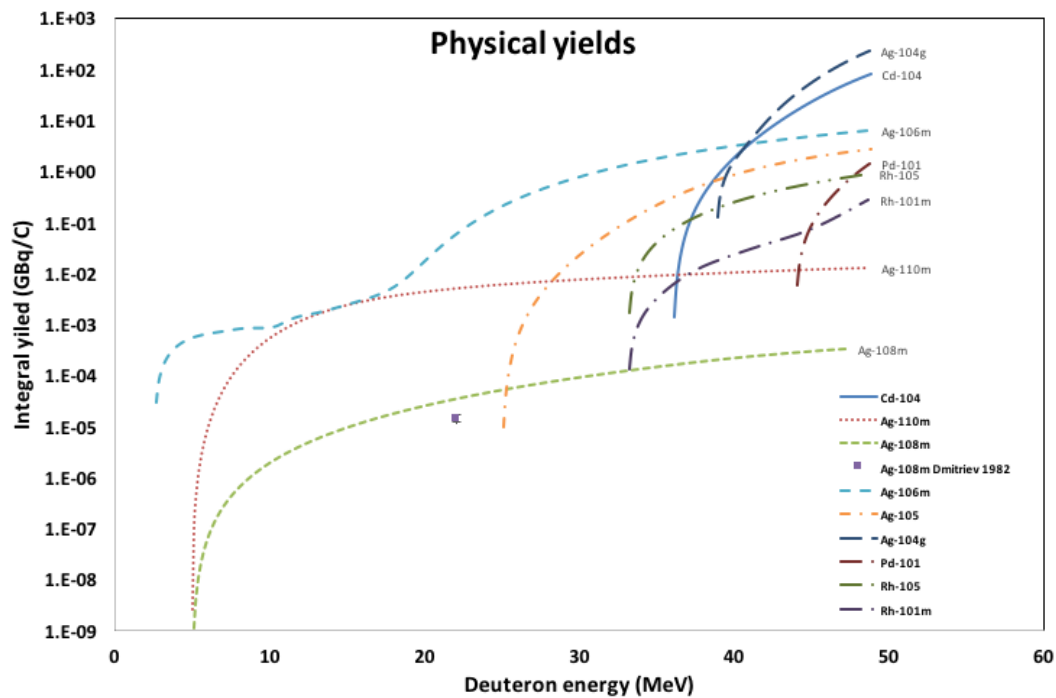

Fig. 11. Integral yields for production of some selected radioisotopes

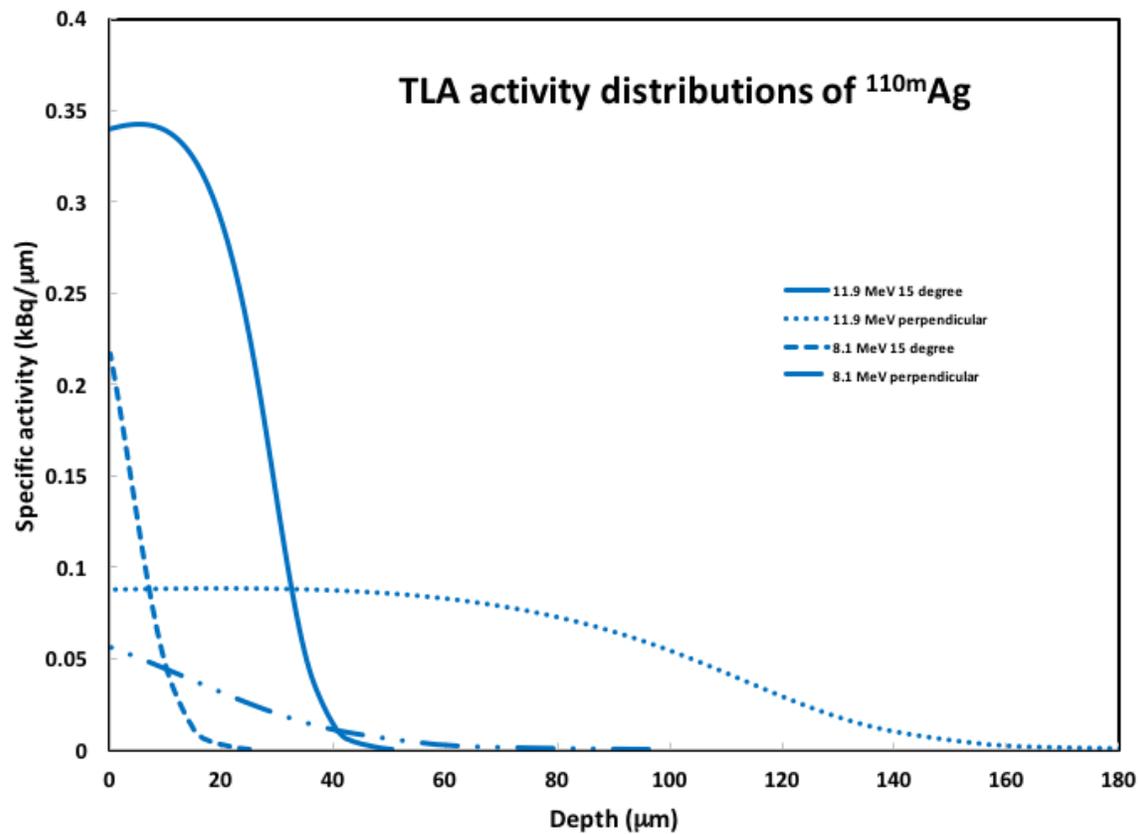

Fig. 12. Specific activity distributions of the $^{110m}$Ag radioisotope (depth profile)

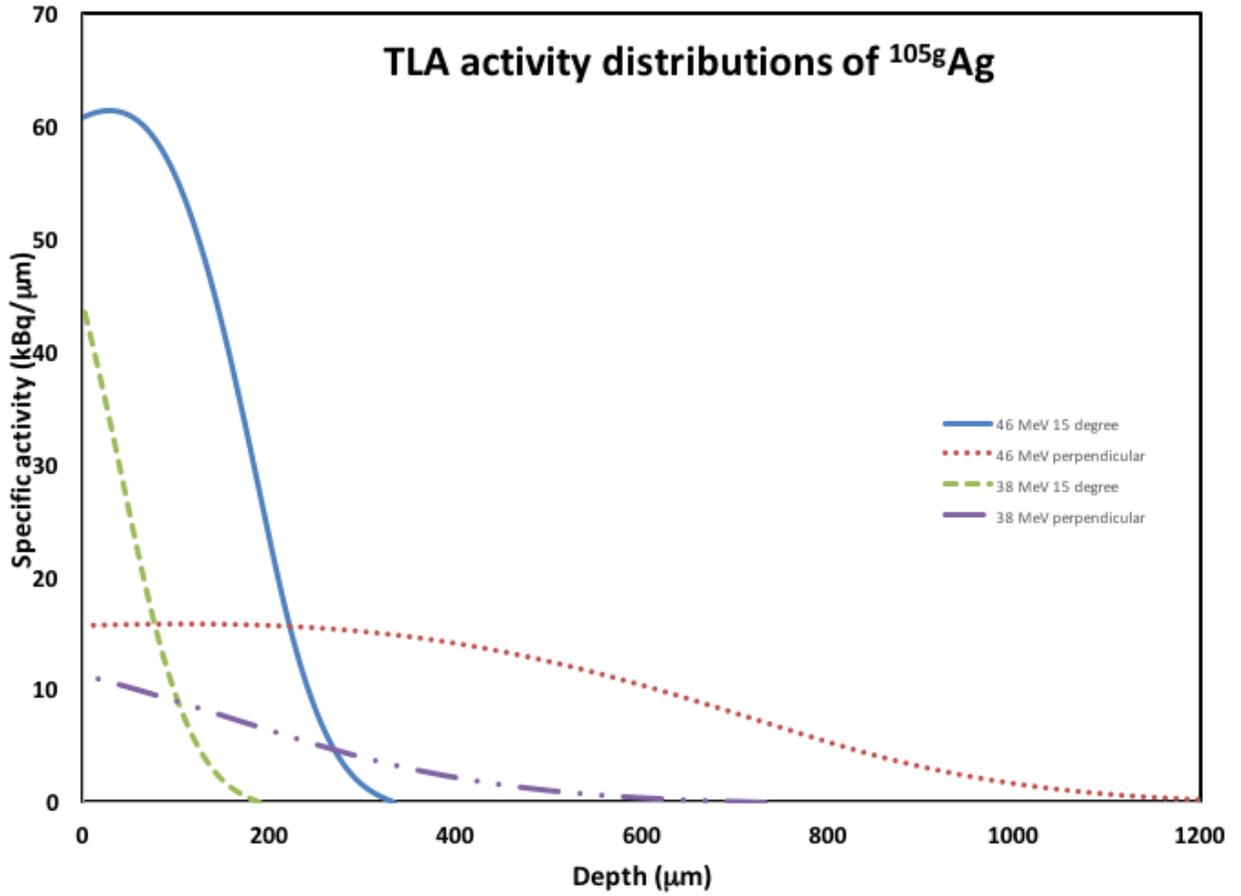

Fig. 13. Specific activity distributions of the $^{105g}$Ag radioisotope (depth profile)